\documentclass[aps,prl,reprint,twocolumn,tightenlines,floatfix,amsmath,amssymb]{revtex4-1}

\usepackage{graphicx}
\usepackage{hyperref}
\usepackage{pdfpages}

\newcommand{\kb}{{k_B}}
\newcommand{\Db}{{\mathcal{D}_B}}
\newcommand{\Tcquasi}{T_{\rm qc}}

\begin{document}

\title{Quasi-condensation in two-dimensional Fermi gases}

\author{Chien-Te Wu}
\author{Brandon M. Anderson}
\author{Rufus Boyack}
\author{K. Levin}
\affiliation{James Franck Institute, University of Chicago, Chicago, Illinois 60637, USA}

\begin{abstract}
In this paper  we follow the analysis and
protocols of recent experiments, combined with simple theory,
to arrive at a physical understanding of quasi-condensation
in two dimensional Fermi gases. We find that quasi-condensation
mirrors Berezinski\u{\i}-Kosterlitz-Thouless behavior in many ways,
including the emergence of a strong zero momentum peak
in the pair momentum distribution.
Importantly, the disappearance of this
quasi-condensate occurs at a reasonably
well defined crossover temperature.
The resulting phase diagram, pair momentum distribution, and algebraic
power law decay are compatible with
recent experiments throughout the continuum from BEC to BCS.
\end{abstract}

\maketitle

Understanding two dimensional (2D) fermionic superfluidity has a long history
relating to the Mermin-Wagner theorem~\cite{MW} and Berezinski\u{\i}~\cite{Berezinskii}, Kosterlitz and Thouless
(BKT) physics~\cite{Kosterlitz}. More
recently it has been viewed as important for addressing
the phase fluctuation picture (and related pseudogap
phenomena) associated with
high-$T_{c}$ superconductors~\cite{Loktev}.
Current interest in 2D bosonic superfluids in ultracold atomic gases
has revealed a general consistency with the
BKT transition~\cite{Cornell,Dalibard,Nist}. 
For 2D fermionic superconductors and superfluids, however, 
it should be emphasized that there is
some historical controversy \cite{Beasley} (beginning with Kosterlitz and Thouless \cite{Kosterlitz})
surrounding observable signatures and applicability of BKT physics.

Thus recent reports \cite{Jochim1,Jochim2} of a form of pair condensation in
2D fermionic gases are particularly exciting. These follow
earlier work addressing the ground state \cite{Turlapov}
and the higher temperature regime, away from condensation \cite{Kohl}.
These experiments \cite{Jochim1,Jochim2} show that
strong normal state pairing is an essential component of 2D 
Fermi superfluids, even in the BCS regime. 
In fact, much of the theory invoked to explain
these experiments was based upon true Bose systems.
A characteristic feature of 2D superfluidity at finite
$T$ is the presence of
narrow peaks in the momentum distribution of the pairs, 
\textit{without} macroscopic occupation of the zero momentum state.
Throughout the paper this will be our definition of
``quasi-condensation.'' 
This quasi-condensation in momentum space is
associated with algebraic decay of coherence in real space. 
Importantly, the BKT-related transition temperature is manifested as
a sudden change in slope of a normalized peak momentum distribution for pairs.

In this paper we present a theory of a 2D Fermi gas near quantum degeneracy
and show how it reproduces rather well the results of these recent experiments \cite{Jochim1,Jochim2} 
through an analysis of
the phase diagram, the pair momentum distribution and algebraic
power laws. Given the ground breaking nature of the experiments, it is
important to have an accompanying theoretical study which
follows exactly the same protocols without any adjustments or phenomenology.
Our approach is to be distinguished from other studies of 2D
Fermi gases
\cite{Petrov,Pathintegral1,Loktev,Pathintegral2,Pathintegral3,Strinati2d,Ohashi2d2015,Ohashi2d2013,Parish2014,Parish2,Bertaina2011,Randeria2d}.
In particular, those addressing BKT physics 
\cite{Petrov,Pathintegral1,Loktev,Pathintegral2,Pathintegral3,Parish2014},
use existing formulae \cite{NelsonKosterlitz,Baym} and
determine the unknown parameters to obtain $T_c^{\rm BKT}$. By
contrast here we reverse the procedure and
follow experimental protocols to thereby
provide a new formula, involving composite bosons,
for the transition temperature
associated with quasi-condensation. In the homogeneous case,
this is analytically
tractable and presented as Eq.~(\ref{eq:tcquasi}) below. 

Importantly, there is a rather abrupt crossover
out of a quasi-condensed phase at a fairly well defined temperature $\Tcquasi$.
In the BEC regime this
matches earlier theoretical
estimates of the BKT transition temperature
which are based on different theoretical
formalisms \cite{Petrov,Pathintegral1,Loktev,Pathintegral2,Pathintegral3}.
We find that $\Tcquasi$ varies continuously with scattering length and, in
reasonable agreement with experiment \cite{Jochim1}, 
the transition appears
at a slightly higher 
temperature for more BCS-like systems. 
We infer that the physics driving this quasi-condensation
derives from implications of the Mermin-Wagner theorem; that is, from
the inability to condense except at zero temperature.
To minimize the free energy, the system remains quasi-condensed for a range
of finite temperatures.
Since we, as in Ref.~\cite{Baym}, make no reference to vortices we cannot argue that our
observations correspond strictly to a BKT scenario \cite{Kosterlitz}, but we can establish
that our findings follow rather precisely those of recent experiments.

\textit{Background theory.$-$}
Theoretical studies of the 2D Fermi gas divide into two
classes: those which build on or extend BCS mean-field theory
\cite{Petrov,Pathintegral1,Loktev,Pathintegral2,Pathintegral3,Randeria2d,Dupuis},
which is the largest class, and those (based on $t$-matrix
schemes) which do not
\cite{Strinati2d,Ohashi2d2015,Ohashi2d2013,Parish2014,Zwerger}.
Here we consider a $t$-matrix theory belonging to the first class.
In the following overview we omit
technical details which can be found in
the Supplemental Material and are extensively discussed elsewhere
\cite{Chen1,ourreview}. There, we also present a comparison with other
theories.

To describe the Fermi gas, we begin by introducing
a pair propagator $\Gamma(Q)$, representing
a Green's function for bosonic, or paired fermionic, degrees of freedom. 
Here we define the vector $Q=\left(i\Omega,\mathbf{q} \right)$, 
where $i\Omega$ is a bosonic Matsubara frequency at temperature $T$
and $\mathbf{q}$ is the pair momentum.
The pair propagator $\Gamma(Q)$ is chosen so that $\Gamma^{-1}(0) = 0$ at a temperature
below a true 3D phase transition temperature (where $\mu_{\rm pair} \equiv 0$)
and, importantly, we impose the condition that this Thouless criterion reproduces
the usual mean-field equation determining the pairing gap, $\Delta \neq 0$. 
We emphasize that $\Delta$ is a pairing gap and not an order parameter. 
In 2D, where $T_c = 0$, this equation at non-zero $T$ is naturally generalized to
$\Gamma^{-1}(0) \propto \mu_{\rm pair}$.
This serves to implement the reasonable assumption that we consider the normal phase for $T>0$ 
without phase coherence, 
but in the presence of a pairing gap so that $\Delta \neq 0$. 

A key component of the theory is the inclusion of fluctuations, or
bosonic degrees of freedom. As we will show,
fluctuations in 2D are necessarily unable to condense, thus guaranteeing
that $\mu_{\rm pair}$ will never vanish for any $T > 0$. 
Because $\Gamma(Q)$ represents a pair propagator, it can be expanded at small $Q$
into the generic form:
\begin{equation}
\Gamma(Q) = \frac {a_0^{-1}}{i \Omega - \Omega_{\mathbf{q}} +\mu_{\rm pair} + i \gamma_Q},
\label{eq:expandt}
\end{equation}
and we associate $\Omega_{\mathbf{q}} \approx {q^2}/{2 M_B}$ 
with a pair dispersion of mass $M_B$.
Throughout we find that we can drop the small lifetime contribution $\gamma_Q$.
Note that
the small $Q$ form of the pair propagator is, up to a constant $a_0$, that of a Bose gas which has no direct inter-boson
interactions, but in which the
bosons interact indirectly via the fermionic medium.

Performing the sum over bosonic Matsubara frequencies $i \Omega$ gives the 
momentum distribution of bosons defined through 
$n_B \left({\mathbf{q}}\right) = a_0 \sum_{i\Omega}\Gamma(Q) = b\left(\Omega_{\mathbf{q}}-\mu_{\rm pair}\right)$, 
where $b(x)=\left(e^{x/\kb T}-1\right)^{-1}$ is the Bose-Einstein distribution function. 
From here it is natural to define a boson number density $n_B$ through
$n_B \equiv \sum_{\mathbf{q}} n_{B}\left({\mathbf{q}}\right) = a_0 \Delta^2$,
where in the second equality we have associated the number density with the 
pairing gap. 
This association is based on the self-energy
and addressed in detail in the Supplemental Material, where we review
the microscopic basis \cite{Chen1,ourreview} of our theory.

\begin{figure*}\centering
\includegraphics[width=7.05in]{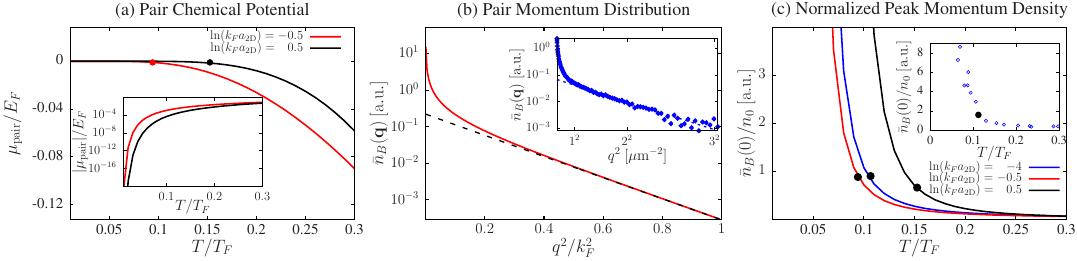}
\caption{(a) Pair chemical potential, $\mu_{\rm pair}$, as a function of temperature at a
scattering length of $\ln (k_Fa_{\mathrm{2D}})=\pm0.5$. 
At high temperatures, the pair chemical potential is large
and negative, whereas it is exponentially suppressed as $T\rightarrow 0$.
The inset shows the same quantity on a logarithmic scale to show the exponentially small
but, non-zero, behavior.
(b) Normalized momentum distribution at $T/T_F=0.07$ and
$\ln (k_Fa_{\mathrm{2D}})=-0.5$.
In the main plot, the red (solid) curve shows the trap-averaged
pair momentum distribution in the high temperature limit. The black
(dashed) curve is the trap-averaged interpolation from a corresponding
Boltzmann distribution. The inset shows the same quantity from Ref.~\cite{Jochim1};
the dashed line is a Boltzmann fit to the experimental data at large $q$.
(c) Peaks in the normalized momentum density as functions of temperature at a range 
of scattering lengths across the BEC-BCS crossover.
The black solid circles are $\Tcquasi$ determined from Eq.~(\ref{eq:tcquasi}).
The insets on the right show an example of corresponding experimental data with $\ln(k_F a_{\rm 2D}) = -0.5$.}
\label{fig:peak}
\end{figure*}

With this formalism we can now determine the 
unknowns that appear in Eq.~(\ref{eq:expandt}).
We use the generalized gap equation
$\Gamma^{-1}(0)= a_0 \mu_{\rm pair}$,
along with the usual BCS equation for determining the fermionic chemical
potential, $\mu$ and the bosonic number equation 
to arrive at
three coupled equations  
\begin{eqnarray}
\sum_{\mathbf{k}}\left[\frac{1-2f\left(E_{\mathbf{k}}\right)}{2E_{\mathbf{k}}}-\frac{1}{2\epsilon_{\mathbf{k}}+\epsilon_{B}}\right]&=&a_{0}\mu_{{\rm pair}}, \label{eq:GP}\\
\label{eq:NB}\sum_{\mathbf{q}}b\left(\frac{q^{2}}{2M_B}-\mu_{{\rm pair}}\right)&=&a_{0}\Delta^{2},\\
\sum_{\mathbf{k}}\left[1-\frac{\xi_{\mathbf{k}}}{E_{\mathbf{k}}}\left(1-2f\left(E_{\mathbf{k}}\right)\right)\right]&=&n,
\end{eqnarray}
where $n$ is the density of fermions.
Here we define the Fermi-Dirac distribution function $f\left(x\right) = \left(e^{x/\kb T} + 1\right)^{-1}$,
the single particle dispersion $\xi_\mathbf{k} = k^2/2m - \mu$
for a fermion of mass $m$, momentum $\mathbf{k}$ and chemical potential $\mu$,
and the Bogoliubov dispersion with gap $\Delta$ which is given by
$E_{\bf k} \equiv \sqrt{\xi_\mathbf{k}^2 + \Delta^2}$.
We have regularized the gap equation in Eq.~(\ref{eq:GP})
by introducing a two particle bound state energy
$\epsilon_{B}=\hbar^{2}/ma_{\mathrm{2D}}^{2}$~\cite{Randeria2d}.
To match with experiment, we use a quasi-2D scattering length 
$a_{\mathrm{2D}}$ parameterized through $\ln(k_F a_{\mathrm{2D}})$. 
We assume throughout that the transverse confinement is sufficient
we can neglect corrections due to a finite transverse trapping length~\cite{[{The quasi-2D scattering length is related to the true 3D scattering length through $a_{\mathrm{2D}}/l_{z}=A\exp\left[-Bl_{z}/a_{\mathrm{3D}}\right]$. (}] [{) Here $l_{z}$ is the transverse confinement length, $A=\sqrt{\pi/0.905}\approx1.8$ and $B=\sqrt{\pi/2}\approx1.2$.}] Petrov2000}.

At $T=0$, we can use the well known solution
$\mu = \epsilon_F - \epsilon_B/2$, $\Delta = \sqrt{2\epsilon_F \epsilon_B}$~\cite{Randeria2d}
along with $\mu_{\rm pair}=0$,
where $\epsilon_F = \pi \hbar^2 n / m$ is the Fermi energy.
At finite temperatures, Eq.~(\ref{eq:NB}) can be inverted exactly to give:
\begin{equation}
\mu_{{\rm pair}} = \kb T\ln\left(1-e^{-n_{B} \lambda_{B}^2}\right),
\label{eq:MUP}
\end{equation}
where $\lambda_B = \sqrt{2\pi \hbar^2 / M_B \kb T}$ is the thermal wavelength for the bosonic pairs.
The pair chemical potential therefore crucially relies on the bosonic phase-space density $\Db = n_{B} \lambda_{B}^2 \sim 1/T$,
which functions roughly as a proxy for inverse temperature.
At low temperature, $\Db \gg 1$ and we find $\mu_{{\rm pair}}/\kb T \sim - e^{-n_{B} \lambda_{B}^2}$, 
or that the chemical potential is exponentially suppressed. 
On the other hand, at high temperatures $\Db \ll 1$ and $\mu_{{\rm pair}} \sim - T \ln T$ which can be substantial. 

\textit{Analysis.$-$} 
Since $\mu_{\rm pair}$ is finite and continuous at all non-zero 
temperatures there can be no true phase transition \cite{DalibardReview}.
Nevertheless there is a rather abrupt threshold from
a moderately large to an exponentially small 
chemical potential. We introduce a tolerance factor $\epsilon$ which
can ultimately be determined from the experimental protocols
\cite{Jochim1}, and which
defines this threshold via the fugacity
$z=e^{\mu_{\rm pair}/\kb T}$. 
The boundary between the low and high temperature behaviors occurs when 
the slope of the fugacity with respect to phase-space density
is of order $\epsilon$:
$dz(\Db)/d\Db \sim \epsilon$.
This will introduce a scale for an effective (BKT-like) crossover temperature:
$\Db = \ln(1/\epsilon)$.

Importantly, there is a very weak (logarithmic) dependence on this tolerance factor
which underlines the fact that the transition will be quite abrupt.
More accessible experimentally \cite{Jochim1} than $\mu_{\rm pair}$
is the behavior of the zero pair momentum peak magnitude, called
$n_B(\mathbf{q}=0) \equiv n_B(0)$. 
The magnitude of this peak is directly related to the
bosonic phase space density through $n_B\left(0\right) = e^{\Db} - 1$.
Thus we can rewrite the crossover constraint on the fugacity as
$dn_B(0)/d\Db \sim  1 / \epsilon$.
In this way we find that a threshold in $\mu_{\rm pair}$ enters as
a slightly rounded knee in $n_B(0)$. Here we use this knee  
to determine an effective
quasi-condensation transition temperature, in much the same way as in experiment. 
In more direct comparisons with Ref. \cite{Jochim1} we find that
$\epsilon$ is approximately one percent. This
corresponds to $\Db \approx 4.6$, which is close to the
Monte Carlo result for the BKT transition 
of a true bosonic gas \cite{MonteCarlo},
which took a typical value of $\Db \approx 4.9$ in Ref.~\cite{Jochim2}.

Converting the transition condition to a form analogous to the 
widely applied \cite{Loktev,Pathintegral1,Pathintegral2,Pathintegral3}
Kosterlitz-Nelson condition \cite{NelsonKosterlitz}, 
$\kb T_c = \frac{\pi}{2} \rho_s(T_c) \hbar^2 / m^2$ for a 
phase stiffness $\rho_s$, we find:
\begin{equation}
\kb \Tcquasi \approx \frac{\pi}{2.3} \frac{\hbar^2 n_B(\Tcquasi)}{M_B(\Tcquasi)}.
\label{eq:tcquasi}
\end{equation}
In the deep-BEC regime (where $n_B/M_B=n/4m$)
this yields $\kb \Tcquasi\approx\frac{1}{9}\epsilon_{F},$
which is similar to estimates in the literature given by 
$k_{B}T_c^{\rm BKT}=\frac{1}{8}\epsilon_{F}$.
Importantly, the present 
expression for $\Tcquasi$ applies 
throughout the BCS-BEC crossover. Towards the BCS limit
the number of bosons decreases, but this is compensated largely
in the crossover temperature by the decrease in bosonic
effective mass.
These analytic arguments apply to a homogeneous system and, following
experimental protocols, they
relate to the characteristics of the non-condensed pairs.

In order to better compare to experiment, we apply the local density approximation (LDA) 
to account for trap effects. 
We note the presence of a trap 
provides only a minor quantitative change 
to the general qualitative picture.
To apply the LDA, we rewrite our equations using the transformations
$\mu\rightarrow\mu\left(\mathbf{R}\right)=\mu_0-\frac{1}{2}m\omega^{2}\mathbf{R}^{2}$,
and $\Delta\rightarrow\Delta\left(\mathbf{R}\right)$, where $\mathbf{R}$ is a local
position, and is not to be confused with the conjugate to $\mathbf{q}$. 
Note that $\mu_0$ is still homogeneous in space (i.e., there is only one degree of freedom)
but that $\Delta\left(\mathbf{R}\right)$ is no longer homogeneous.
We now have $\xi_{\mathbf{k}}\rightarrow\xi_{\mathbf{k}}\left(\mathbf{R}\right)=k^2/2m-\mu\left(\mathbf{R}\right)$,
and $E_{\mathbf{k}}\rightarrow E_{\mathbf{k}}\left(\mathbf{R}\right)=\sqrt{\xi^{2}_{\mathbf{k}}\left(\mathbf{R}\right)+\Delta^{2}\left(\mathbf{R}\right)}$.
Similarly, $M_B\rightarrow M_B\left(\mathbf{R}\right)$,
$n_B\rightarrow n_B\left(\mathbf{R}\right)$, $a_{0}\rightarrow a_{0}\left(\mathbf{R}\right)$, etc., 
through these same substitutions. We also define a trap-integrated 
momentum distribution: 
$\bar{n}_B(\mathbf{k}) = \int n_B(\mathbf{k},\mathbf{R}) d^2\mathbf{R} $.
For details of the LDA, including the parameters used, see the Supplemental Material.

\textit{Comparison with experiment.$-$} 
We now compare our theory with the recent experimental results in Refs.~\cite{Jochim1,Jochim2},
using our numerical results for the trapped case
as ``data'' analogous to the experiment. In order to probe the 
momentum distribution of bosonic pairs at low temperatures, in 
Fig.~\ref{fig:peak}(a)
we plot the pair chemical potential versus temperature
for two values of $\mathrm{ln}(k_Fa_{\mathrm{2D}})=\pm 0.5$, along with 
an enlarged plot of
$|\mu_{\rm pair}(T)|$ which is presented in the inset. 
The dots illustrate the crossover points 
associated with the threshold discussed earlier; here $\mu_{\rm pair}$ begins 
to appreciably deviate from zero, thus marking the transition out of the quasi-condensed state. 
Figure \ref{fig:peak}(b) shows an example of the trap-integrated 
pair momentum distribution $\bar{n}_B(\mathbf{q})$ at $T<\Tcquasi$. 
The small chemical potential $\mu_{\rm pair}$ results in a sharply peaked distribution $\bar{n}_B(\mathbf{q})$
as $\mathbf{q}\rightarrow 0$. This behavior is similar to the results observed in experiment~\cite{Jochim1},
as shown in the inset. 
Therefore this peak, or signature of quasi-condensation, emerges when the
pair chemical potential becomes sufficiently small.

\begin{figure}
\includegraphics[width=3.4in]{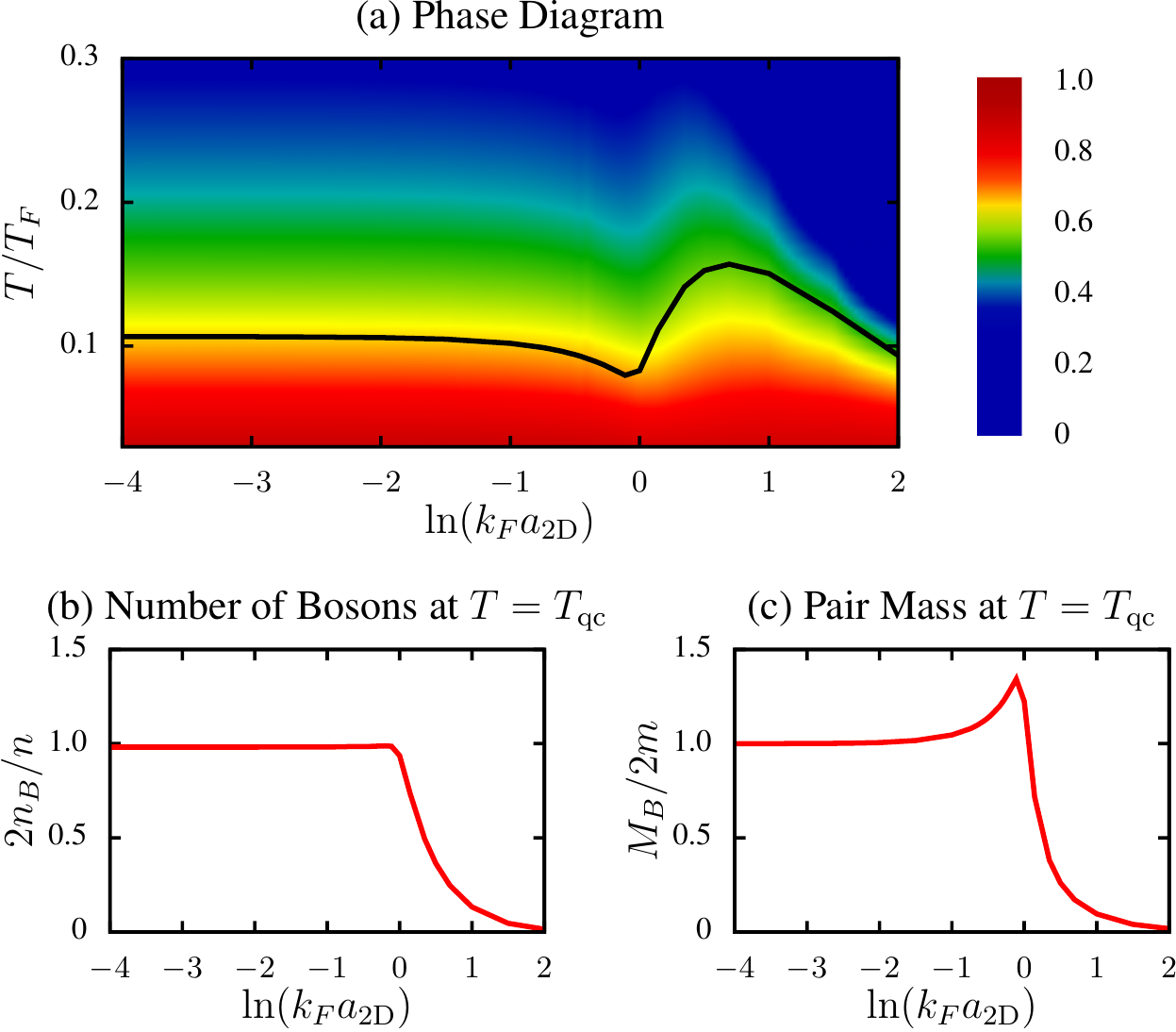}
\caption{(a) The phase diagram of a quasi-condensation crossover in a 2D fermi gas
as a function of the scattering length $\ln(k_F a_{\rm 2D})$.
The black curve in the phase diagram represents $\Tcquasi$,
and the colored shading
represents the non-thermal fraction (see text and Fig.~\ref{fig:peak}(b)).
The main structure can be understood from Eq.~(\ref{eq:tcquasi}), along with
the pair mass (b) $M_B(\Tcquasi)$ and bosonic number density (c) $n_B(\Tcquasi)$ at $\Tcquasi$.
On the BEC side, $\Tcquasi$ limits to a constant value.
A maximum of the pair mass near $\ln(k_F a_{\rm 2D}) = 0$ results in a drop in $\Tcquasi$ at the same point.
In the BCS limit the transition temperature drops as the bosonic number density declines more quickly than the pair mass.
All quantities are calculated from ${R}=0$ data in the LDA.
}
\label{fig:phase}
\end{figure}

To quantify an effective crossover transition temperature, we focus on
the ratio of the peak magnitude $\bar{n}_B(0)$ of this momentum distribution normalized to the peak 
number density in the center of the trap, $n_0=n(\mathbf{R}=0)$, following the experimental protocol \cite{Jochim1}. 
This is plotted for three different values of $\ln\left(k_Fa_{\mathrm{2D}}\right)$ in
Fig.~\ref{fig:peak}(c)
with the dots indicating the knee
(assuming a one percent tolerance factor). 
This allows us to arrive at a BKT-like transition temperature in a trap
as a function of scattering length.
The inset plots the experimental results for comparison.
For our ``data" the peak of the
momentum distribution grows exponentially at low temperatures. In contrast, the experimental data does not grow
quickly enough to distinguish between exponential growth and a sharp peak in this distribution.
More generally, we
note that the deviations between experiment and our theory consistently 
suggest that the absolute value of our chemical potential is too small.
This, in turn, reflects the behavior of the
BCS-like gap equation which sets the scale for $\mu_{\rm pair}$
through Eq.~(\ref{eq:GP}).

We next use this analysis to obtain the phase diagram as a function
of interaction strengths investigated in experiment. 
The results are shown in Fig.~\ref{fig:phase}(a) plotted against $\mathrm{ln}(k_Fa_{\rm 2D})$. 
We can associate a BCS-like phase with positive fermionic 
$\mu$ which appears when $\ln (k_Fa_{\mathrm{2D}}) > 0$.
Figures \ref{fig:phase}(b) and \ref{fig:phase}(c) indicate the numerator ($n_B$) and denominator ($M_B$)
components of $\Tcquasi$ as shown in Eq.~(\ref{eq:tcquasi}) as a function of scattering length. 
The color coding indicates the non-thermal fraction which is found from Fig.~\ref{fig:peak}(b)
as the area between the momentum distribution (solid curve) and its high temperature asymptote (dashed line).
On the BEC side, we find that $\Tcquasi$ saturates. As the scattering length is increased, the transition
temperature begins to drop before rising to a local maximum and 
then falling off in the deep BCS regime. 
While the values are rather similar, this non-monotonic behavior 
is not as directly seen in experiment, although it is suggested in their plots of the non-thermal
fraction. 
It should also be noted that our theory for the transition temperature is valid across the entire BEC-BCS spectrum,
rather than as two endpoint cases as often studied \cite{Petrov,Parish2014}.

\begin{figure}
\includegraphics[width=3.3in]{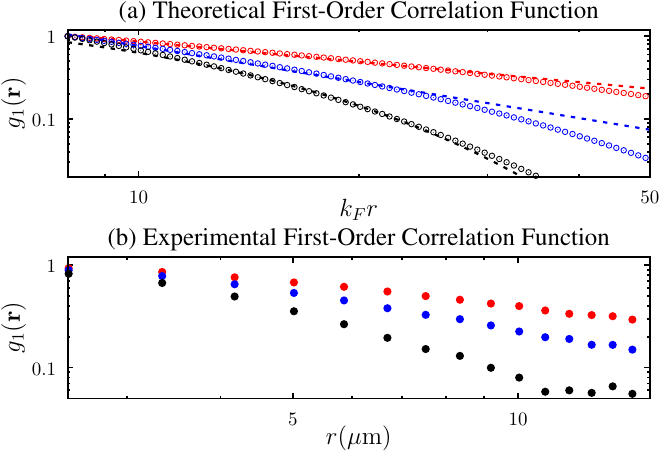}
\caption{Correlation function, $g_1(\mathbf{r})$,
at large distances found through a Fourier transform of the trap-averaged momentum-density
in the LDA approximation. Both the theoretical and experimental results
are presented on a log-log scale.
In the top panel, we consider three different temperatures at a fixed
scattering length $\ln (k_Fa_{\mathrm{2D}})=-0.5$, where $\Tcquasi=0.094 T_F$.
From top to bottom, $T/T_F=0.06$, $T/T_F=0.09$, and $T/T_F=0.12$. For $T<\Tcquasi$,
we fit to a power law expression (dashed lines). The dashed curve
for $T/T_F=0.12$ is an exponential fit. The bottom panel is the experimental comparison.
The temperatures for the theory curves are close to those considered in 
experiment.}
\label{fig:g1}
\end{figure}

Finally, in Fig.~\ref{fig:g1} we present
the correlation function $g_1(\mathbf{r})$ determined
from the Fourier transform of the trap integrated momentum distribution $\bar{n}_B(\mathbf{q})$, 
again following the experimental protocol \cite{Jochim2}. 
We fit to a power law for an intermediate range of $r$ corresponding roughly to that used in the experimental data~\cite{Jochim2}.
It should be noted, however, that the power law regime appears
slightly more extended in experiment than in theory. With our analytic insight we believe
there may be better fits to our ``data" by
considering the structure of the momentum distribution
at $\mathbf{q}\rightarrow 0$ \cite{fitwords}.
Nevertheless following experiment, we
find a reasonable fit to a power law in this range at low temperature, $g_1(\mathbf{r}) \sim 1/r^\eta$, and a crossover to an exponential fit at higher temperatures, $g_1(\mathbf{r}) \sim e^{-r/\xi}$.
Our power laws lie in the range of $0.8<\eta<1.45$. These values are close to the power laws observed in the experiment of $0.6<\eta<1.4$, 
which appear universal near the expected BKT transition temperature, yet are far from the predicted scaling of $\eta \leq 1/4$.

\textit{Conclusions.$-$}
The favorable comparisons between theory and experiment in Figs.~\ref{fig:peak}$-$\ref{fig:g1}
provide helpful insights into the behavior of
2D Fermi gases. 
Central to our picture,
is the relation between the zero momentum peak in the
pair distribution function and the
small pair chemical potential $\mu_{\rm pair}$.
As consistent with the Mermin-Wagner theorem, $\mu_{\rm pair}$ is shown
to never vanish except at zero temperature. 
We argue that it is this inability to fully condense which ultimately drives
quasi-condensation.
Importantly,
with increasing temperature there is a rather abrupt
transition from this quasi-condensed phase which, following
experimental protocols \cite{Jochim1,Jochim2} is
reflective of BKT physics.

Our approach should be contrasted with
contributions to the theoretical
literature which address BKT physics
\cite{Petrov,Pathintegral1,Loktev,Pathintegral2,Pathintegral3,Parish2014},
by solving for the phase stiffness, etc. parameters that appear in
the usual formulae \cite{Baym,NelsonKosterlitz} for the BKT transition temperature.
Here we reverse the logic
in order to present an alternative viewpoint.
We follow the experimental procedure to thereby
provide a new formula 
(see Eq.~(\ref{eq:tcquasi}) in the absence of a trap)
for the transition temperature
associated with quasi-condensation. This 
can be expressed in terms of the bosonic phase space
density, $\Db$, which we find, as in other literature
\cite{MonteCarlo}, 
is around four; as expected \cite{DalibardReview}, 
$\Db$ exhibits no sharp transition.
We stress that our expression is associated with composite
bosons whose mass and number density
vary significantly and continuously from BCS to BEC.

\acknowledgments{
\textit{Acknowledgments.$-$}
This work was supported by NSF-DMR-MRSEC 1420709.
We are particularly grateful to Selim Jochim, Martin Ries,
and Puneet Murthy for sharing their data and for 
helpful conversations regarding their experiment and feedback.
We also thank Colin Parker for enlightening discussions.}

\bibliography{Review}

\begin{thebibliography}{33}%
\makeatletter
\providecommand \@ifxundefined [1]{%
 \@ifx{#1\undefined}
}%
\providecommand \@ifnum [1]{%
 \ifnum #1\expandafter \@firstoftwo
 \else \expandafter \@secondoftwo
 \fi
}%
\providecommand \@ifx [1]{%
 \ifx #1\expandafter \@firstoftwo
 \else \expandafter \@secondoftwo
 \fi
}%
\providecommand \natexlab [1]{#1}%
\providecommand \enquote  [1]{``#1''}%
\providecommand \bibnamefont  [1]{#1}%
\providecommand \bibfnamefont [1]{#1}%
\providecommand \citenamefont [1]{#1}%
\providecommand \href@noop [0]{\@secondoftwo}%
\providecommand \href [0]{\begingroup \@sanitize@url \@href}%
\providecommand \@href[1]{\@@startlink{#1}\@@href}%
\providecommand \@@href[1]{\endgroup#1\@@endlink}%
\providecommand \@sanitize@url [0]{\catcode `\\12\catcode `\$12\catcode
  `\&12\catcode `\#12\catcode `\^12\catcode `\_12\catcode `\%12\relax}%
\providecommand \@@startlink[1]{}%
\providecommand \@@endlink[0]{}%
\providecommand \url  [0]{\begingroup\@sanitize@url \@url }%
\providecommand \@url [1]{\endgroup\@href {#1}{\urlprefix }}%
\providecommand \urlprefix  [0]{URL }%
\providecommand \Eprint [0]{\href }%
\providecommand \doibase [0]{http://dx.doi.org/}%
\providecommand \selectlanguage [0]{\@gobble}%
\providecommand \bibinfo  [0]{\@secondoftwo}%
\providecommand \bibfield  [0]{\@secondoftwo}%
\providecommand \translation [1]{[#1]}%
\providecommand \BibitemOpen [0]{}%
\providecommand \bibitemStop [0]{}%
\providecommand \bibitemNoStop [0]{.\EOS\space}%
\providecommand \EOS [0]{\spacefactor3000\relax}%
\providecommand \BibitemShut  [1]{\csname bibitem#1\endcsname}%
\let\auto@bib@innerbib\@empty
\bibitem [{\citenamefont {Mermin}\ and\ \citenamefont {Wagner}(1966)}]{MW}%
  \BibitemOpen
  \bibfield  {author} {\bibinfo {author} {\bibfnamefont {N.~D.}\ \bibnamefont
  {Mermin}}\ and\ \bibinfo {author} {\bibfnamefont {H.}~\bibnamefont
  {Wagner}},\ }\href {\doibase 10.1103/PhysRevLett.17.1133} {\bibfield
  {journal} {\bibinfo  {journal} {Phys. Rev. Lett.}\ }\textbf {\bibinfo
  {volume} {17}},\ \bibinfo {pages} {1133} (\bibinfo {year}
  {1966})}\BibitemShut {NoStop}%
\bibitem [{\citenamefont {Berezinski\u{\i}}(1972)}]{Berezinskii}%
  \BibitemOpen
  \bibfield  {author} {\bibinfo {author} {\bibfnamefont {V.}~\bibnamefont
  {Berezinski\u{\i}}},\ }\href@noop {} {\bibfield  {journal} {\bibinfo
  {journal} {Sov. Phys. JETP}\ }\textbf {\bibinfo {volume} {34}},\ \bibinfo
  {pages} {610} (\bibinfo {year} {1972})}\BibitemShut {NoStop}%
\bibitem [{\citenamefont {Kosterlitz}\ and\ \citenamefont
  {Thouless}(1973)}]{Kosterlitz}%
  \BibitemOpen
  \bibfield  {author} {\bibinfo {author} {\bibfnamefont {J.~M.}\ \bibnamefont
  {Kosterlitz}}\ and\ \bibinfo {author} {\bibfnamefont {D.~J.}\ \bibnamefont
  {Thouless}},\ }\href@noop {} {\bibfield  {journal} {\bibinfo  {journal} {J.
  Phys. C. Solid State}\ }\textbf {\bibinfo {volume} {6}},\ \bibinfo {pages}
  {1181} (\bibinfo {year} {1973})}\BibitemShut {NoStop}%
\bibitem [{\citenamefont {Loktev}\ \emph {et~al.}(2001)\citenamefont {Loktev},
  \citenamefont {Quick},\ and\ \citenamefont {Sharapov}}]{Loktev}%
  \BibitemOpen
  \bibfield  {author} {\bibinfo {author} {\bibfnamefont {V.~M.}\ \bibnamefont
  {Loktev}}, \bibinfo {author} {\bibfnamefont {R.~M.}\ \bibnamefont {Quick}}, \
  and\ \bibinfo {author} {\bibfnamefont {S.~G.}\ \bibnamefont {Sharapov}},\
  }\href@noop {} {\bibfield  {journal} {\bibinfo  {journal} {Physics Reports}\
  }\textbf {\bibinfo {volume} {349}},\ \bibinfo {pages} {1} (\bibinfo {year}
  {2001})}\BibitemShut {NoStop}%
\bibitem [{\citenamefont {Tung}\ \emph {et~al.}(2010)\citenamefont {Tung},
  \citenamefont {Lamporesi}, \citenamefont {Lobser}, \citenamefont {Xia},\ and\
  \citenamefont {Cornell}}]{Cornell}%
  \BibitemOpen
  \bibfield  {author} {\bibinfo {author} {\bibfnamefont {S.}~\bibnamefont
  {Tung}}, \bibinfo {author} {\bibfnamefont {G.}~\bibnamefont {Lamporesi}},
  \bibinfo {author} {\bibfnamefont {D.}~\bibnamefont {Lobser}}, \bibinfo
  {author} {\bibfnamefont {L.}~\bibnamefont {Xia}}, \ and\ \bibinfo {author}
  {\bibfnamefont {E.~A.}\ \bibnamefont {Cornell}},\ }\href {\doibase
  10.1103/PhysRevLett.105.230408} {\bibfield  {journal} {\bibinfo  {journal}
  {Phys. Rev. Lett.}\ }\textbf {\bibinfo {volume} {105}},\ \bibinfo {pages}
  {230408} (\bibinfo {year} {2010})}\BibitemShut {NoStop}%
\bibitem [{\citenamefont {Hadzibabic}\ \emph {et~al.}(2006)\citenamefont
  {Hadzibabic}, \citenamefont {Kruger}, \citenamefont {Cheneau}, \citenamefont
  {Battelier},\ and\ \citenamefont {Dalibard}}]{Dalibard}%
  \BibitemOpen
  \bibfield  {author} {\bibinfo {author} {\bibfnamefont {A.}~\bibnamefont
  {Hadzibabic}}, \bibinfo {author} {\bibfnamefont {P.}~\bibnamefont {Kruger}},
  \bibinfo {author} {\bibfnamefont {M.}~\bibnamefont {Cheneau}}, \bibinfo
  {author} {\bibfnamefont {B.}~\bibnamefont {Battelier}}, \ and\ \bibinfo
  {author} {\bibfnamefont {J.}~\bibnamefont {Dalibard}},\ }\href@noop {}
  {\bibfield  {journal} {\bibinfo  {journal} {Nature}\ }\textbf {\bibinfo
  {volume} {441}},\ \bibinfo {pages} {1118} (\bibinfo {year}
  {2006})}\BibitemShut {NoStop}%
\bibitem [{\citenamefont {Clad\'e}\ \emph {et~al.}(2009)\citenamefont
  {Clad\'e}, \citenamefont {Ryu}, \citenamefont {Ramanathan}, \citenamefont
  {Helmerson},\ and\ \citenamefont {Phillips}}]{Nist}%
  \BibitemOpen
  \bibfield  {author} {\bibinfo {author} {\bibfnamefont {P.}~\bibnamefont
  {Clad\'e}}, \bibinfo {author} {\bibfnamefont {C.}~\bibnamefont {Ryu}},
  \bibinfo {author} {\bibfnamefont {A.}~\bibnamefont {Ramanathan}}, \bibinfo
  {author} {\bibfnamefont {K.}~\bibnamefont {Helmerson}}, \ and\ \bibinfo
  {author} {\bibfnamefont {W.~D.}\ \bibnamefont {Phillips}},\ }\href {\doibase
  10.1103/PhysRevLett.102.170401} {\bibfield  {journal} {\bibinfo  {journal}
  {Phys. Rev. Lett.}\ }\textbf {\bibinfo {volume} {102}},\ \bibinfo {pages}
  {170401} (\bibinfo {year} {2009})}\BibitemShut {NoStop}%
\bibitem [{\citenamefont {Beasley}\ \emph {et~al.}(1979)\citenamefont
  {Beasley}, \citenamefont {Mooij},\ and\ \citenamefont {Orlando}}]{Beasley}%
  \BibitemOpen
  \bibfield  {author} {\bibinfo {author} {\bibfnamefont {M.~R.}\ \bibnamefont
  {Beasley}}, \bibinfo {author} {\bibfnamefont {J.~E.}\ \bibnamefont {Mooij}},
  \ and\ \bibinfo {author} {\bibfnamefont {T.~P.}\ \bibnamefont {Orlando}},\
  }\href {\doibase 10.1103/PhysRevLett.42.1165} {\bibfield  {journal} {\bibinfo
   {journal} {Phys. Rev. Lett.}\ }\textbf {\bibinfo {volume} {42}},\ \bibinfo
  {pages} {1165} (\bibinfo {year} {1979})}\BibitemShut {NoStop}%
\bibitem [{\citenamefont {Ries}\ \emph {et~al.}(2015)\citenamefont {Ries},
  \citenamefont {Wenz}, \citenamefont {Z\"urn}, \citenamefont {Bayha},
  \citenamefont {Boettcher}, \citenamefont {Kedar}, \citenamefont {Murthy},
  \citenamefont {Neidig}, \citenamefont {Lompe},\ and\ \citenamefont
  {Jochim}}]{Jochim1}%
  \BibitemOpen
  \bibfield  {author} {\bibinfo {author} {\bibfnamefont {M.~G.}\ \bibnamefont
  {Ries}}, \bibinfo {author} {\bibfnamefont {A.~N.}\ \bibnamefont {Wenz}},
  \bibinfo {author} {\bibfnamefont {G.}~\bibnamefont {Z\"urn}}, \bibinfo
  {author} {\bibfnamefont {L.}~\bibnamefont {Bayha}}, \bibinfo {author}
  {\bibfnamefont {I.}~\bibnamefont {Boettcher}}, \bibinfo {author}
  {\bibfnamefont {D.}~\bibnamefont {Kedar}}, \bibinfo {author} {\bibfnamefont
  {P.~A.}\ \bibnamefont {Murthy}}, \bibinfo {author} {\bibfnamefont
  {M.}~\bibnamefont {Neidig}}, \bibinfo {author} {\bibfnamefont
  {T.}~\bibnamefont {Lompe}}, \ and\ \bibinfo {author} {\bibfnamefont
  {S.}~\bibnamefont {Jochim}},\ }\href {\doibase
  10.1103/PhysRevLett.114.230401} {\bibfield  {journal} {\bibinfo  {journal}
  {Phys. Rev. Lett.}\ }\textbf {\bibinfo {volume} {114}},\ \bibinfo {pages}
  {230401} (\bibinfo {year} {2015})}\BibitemShut {NoStop}%
\bibitem [{\citenamefont {Murthy}\ \emph {et~al.}(2015)\citenamefont {Murthy},
  \citenamefont {Boettcher}, \citenamefont {Bayha}, \citenamefont {Holzmann},
  \citenamefont {Kedar}, \citenamefont {Neidig}, \citenamefont {Ries},
  \citenamefont {Wenz}, \citenamefont {Z\"{u}rn},\ and\ \citenamefont
  {Jochim}}]{Jochim2}%
  \BibitemOpen
  \bibfield  {author} {\bibinfo {author} {\bibfnamefont {P.~A.}\ \bibnamefont
  {Murthy}}, \bibinfo {author} {\bibfnamefont {I.}~\bibnamefont {Boettcher}},
  \bibinfo {author} {\bibfnamefont {L.}~\bibnamefont {Bayha}}, \bibinfo
  {author} {\bibfnamefont {M.}~\bibnamefont {Holzmann}}, \bibinfo {author}
  {\bibfnamefont {D.}~\bibnamefont {Kedar}}, \bibinfo {author} {\bibfnamefont
  {M.}~\bibnamefont {Neidig}}, \bibinfo {author} {\bibfnamefont {M.~G.}\
  \bibnamefont {Ries}}, \bibinfo {author} {\bibfnamefont {A.~N.}\ \bibnamefont
  {Wenz}}, \bibinfo {author} {\bibfnamefont {G.}~\bibnamefont {Z\"{u}rn}}, \
  and\ \bibinfo {author} {\bibfnamefont {S.}~\bibnamefont {Jochim}},\ }\href
  {\doibase 10.1103/PhysRevLett.115.010401} {\bibfield  {journal} {\bibinfo
  {journal} {Phys. Rev. Lett.}\ }\textbf {\bibinfo {volume} {115}},\ \bibinfo
  {pages} {10401} (\bibinfo {year} {2015})}\BibitemShut {NoStop}%
\bibitem [{\citenamefont {Makhalov}\ \emph {et~al.}(2014)\citenamefont
  {Makhalov}, \citenamefont {Martiyanov},\ and\ \citenamefont
  {Turlapov}}]{Turlapov}%
  \BibitemOpen
  \bibfield  {author} {\bibinfo {author} {\bibfnamefont {V.}~\bibnamefont
  {Makhalov}}, \bibinfo {author} {\bibfnamefont {K.}~\bibnamefont
  {Martiyanov}}, \ and\ \bibinfo {author} {\bibfnamefont {A.}~\bibnamefont
  {Turlapov}},\ }\href {\doibase 10.1103/PhysRevLett.112.045301} {\bibfield
  {journal} {\bibinfo  {journal} {Phys. Rev. Lett.}\ }\textbf {\bibinfo
  {volume} {112}},\ \bibinfo {pages} {1} (\bibinfo {year} {2014})}\BibitemShut
  {NoStop}%
\bibitem [{\citenamefont {Feld}\ \emph {et~al.}(2011)\citenamefont {Feld},
  \citenamefont {Frohlich}, \citenamefont {Vogt}, \citenamefont {M},\ and\
  \citenamefont {Kohl}}]{Kohl}%
  \BibitemOpen
  \bibfield  {author} {\bibinfo {author} {\bibfnamefont {M.}~\bibnamefont
  {Feld}}, \bibinfo {author} {\bibfnamefont {B.}~\bibnamefont {Frohlich}},
  \bibinfo {author} {\bibfnamefont {E.}~\bibnamefont {Vogt}}, \bibinfo {author}
  {\bibfnamefont {K.}~\bibnamefont {M}}, \ and\ \bibinfo {author}
  {\bibfnamefont {M.}~\bibnamefont {Kohl}},\ }\href@noop {} {\bibfield
  {journal} {\bibinfo  {journal} {Nature}\ }\textbf {\bibinfo {volume} {480}},\
  \bibinfo {pages} {75} (\bibinfo {year} {2011})}\BibitemShut {NoStop}%
\bibitem [{\citenamefont {Petrov}\ \emph {et~al.}(2003)\citenamefont {Petrov},
  \citenamefont {Baranov},\ and\ \citenamefont {Shlyapnikov}}]{Petrov}%
  \BibitemOpen
  \bibfield  {author} {\bibinfo {author} {\bibfnamefont {D.~S.}\ \bibnamefont
  {Petrov}}, \bibinfo {author} {\bibfnamefont {M.~A.}\ \bibnamefont {Baranov}},
  \ and\ \bibinfo {author} {\bibfnamefont {G.~V.}\ \bibnamefont
  {Shlyapnikov}},\ }\href {\doibase 10.1103/PhysRevA.67.031601} {\bibfield
  {journal} {\bibinfo  {journal} {Phys. Rev. A}\ }\textbf {\bibinfo {volume}
  {67}},\ \bibinfo {pages} {031601} (\bibinfo {year} {2003})}\BibitemShut
  {NoStop}%
\bibitem [{\citenamefont {Babaev}\ and\ \citenamefont
  {Kleinert}(1999)}]{Pathintegral1}%
  \BibitemOpen
  \bibfield  {author} {\bibinfo {author} {\bibfnamefont {E.}~\bibnamefont
  {Babaev}}\ and\ \bibinfo {author} {\bibfnamefont {H.}~\bibnamefont
  {Kleinert}},\ }\href {\doibase 10.1103/PhysRevB.59.12083} {\bibfield
  {journal} {\bibinfo  {journal} {Phys. Rev. B}\ }\textbf {\bibinfo {volume}
  {59}},\ \bibinfo {pages} {12083} (\bibinfo {year} {1999})}\BibitemShut
  {NoStop}%
\bibitem [{\citenamefont {Botelho}\ and\ \citenamefont {S\'a~de
  Melo}(2006)}]{Pathintegral2}%
  \BibitemOpen
  \bibfield  {author} {\bibinfo {author} {\bibfnamefont {S.~S.}\ \bibnamefont
  {Botelho}}\ and\ \bibinfo {author} {\bibfnamefont {C.~A.~R.}\ \bibnamefont
  {S\'a~de Melo}},\ }\href {\doibase 10.1103/PhysRevLett.96.040404} {\bibfield
  {journal} {\bibinfo  {journal} {Phys. Rev. Lett.}\ }\textbf {\bibinfo
  {volume} {96}},\ \bibinfo {pages} {040404} (\bibinfo {year}
  {2006})}\BibitemShut {NoStop}%
\bibitem [{\citenamefont {Salasnich}\ \emph {et~al.}(2013)\citenamefont
  {Salasnich}, \citenamefont {Marchetti},\ and\ \citenamefont
  {Toigo}}]{Pathintegral3}%
  \BibitemOpen
  \bibfield  {author} {\bibinfo {author} {\bibfnamefont {L.}~\bibnamefont
  {Salasnich}}, \bibinfo {author} {\bibfnamefont {P.~A.}\ \bibnamefont
  {Marchetti}}, \ and\ \bibinfo {author} {\bibfnamefont {F.}~\bibnamefont
  {Toigo}},\ }\href {\doibase 10.1103/PhysRevA.88.053612} {\bibfield  {journal}
  {\bibinfo  {journal} {Phys. Rev. A}\ }\textbf {\bibinfo {volume} {88}},\
  \bibinfo {pages} {053612} (\bibinfo {year} {2013})}\BibitemShut {NoStop}%
\bibitem [{\citenamefont {Marsiglio}\ \emph {et~al.}(2015)\citenamefont
  {Marsiglio}, \citenamefont {Pieri}, \citenamefont {Perali}, \citenamefont
  {Palestini},\ and\ \citenamefont {Strinati}}]{Strinati2d}%
  \BibitemOpen
  \bibfield  {author} {\bibinfo {author} {\bibfnamefont {F.}~\bibnamefont
  {Marsiglio}}, \bibinfo {author} {\bibfnamefont {P.}~\bibnamefont {Pieri}},
  \bibinfo {author} {\bibfnamefont {A.}~\bibnamefont {Perali}}, \bibinfo
  {author} {\bibfnamefont {F.}~\bibnamefont {Palestini}}, \ and\ \bibinfo
  {author} {\bibfnamefont {G.~C.}\ \bibnamefont {Strinati}},\ }\href {\doibase
  10.1103/PhysRevB.91.054509} {\bibfield  {journal} {\bibinfo  {journal} {Phys.
  Rev. B}\ }\textbf {\bibinfo {volume} {91}},\ \bibinfo {pages} {054509}
  (\bibinfo {year} {2015})}\BibitemShut {NoStop}%
\bibitem [{\citenamefont {Matsumoto}\ \emph {et~al.}(2015)\citenamefont
  {Matsumoto}, \citenamefont {Inotani},\ and\ \citenamefont
  {Ohashi}}]{Ohashi2d2015}%
  \BibitemOpen
  \bibfield  {author} {\bibinfo {author} {\bibfnamefont {M.}~\bibnamefont
  {Matsumoto}}, \bibinfo {author} {\bibfnamefont {D.}~\bibnamefont {Inotani}},
  \ and\ \bibinfo {author} {\bibfnamefont {Y.}~\bibnamefont {Ohashi}},\ }\href
  {http://arxiv.org/abs/1507.05149} {\  (\bibinfo {year} {2015})},\ \Eprint
  {http://arxiv.org/abs/1507.05149} {arXiv:1507.05149} \BibitemShut {NoStop}%
\bibitem [{\citenamefont {Watanabe}\ \emph {et~al.}(2013)\citenamefont
  {Watanabe}, \citenamefont {Tsuchiya},\ and\ \citenamefont
  {Ohashi}}]{Ohashi2d2013}%
  \BibitemOpen
  \bibfield  {author} {\bibinfo {author} {\bibfnamefont {R.}~\bibnamefont
  {Watanabe}}, \bibinfo {author} {\bibfnamefont {S.}~\bibnamefont {Tsuchiya}},
  \ and\ \bibinfo {author} {\bibfnamefont {Y.}~\bibnamefont {Ohashi}},\ }\href
  {\doibase 10.1103/PhysRevA.88.013637} {\bibfield  {journal} {\bibinfo
  {journal} {Phys. Rev. A}\ }\textbf {\bibinfo {volume} {88}},\ \bibinfo
  {pages} {013637} (\bibinfo {year} {2013})}\BibitemShut {NoStop}%
\bibitem [{\citenamefont {Bauer}\ \emph {et~al.}(2014)\citenamefont {Bauer},
  \citenamefont {Parish},\ and\ \citenamefont {Enss}}]{Parish2014}%
  \BibitemOpen
  \bibfield  {author} {\bibinfo {author} {\bibfnamefont {M.}~\bibnamefont
  {Bauer}}, \bibinfo {author} {\bibfnamefont {M.~M.}\ \bibnamefont {Parish}}, \
  and\ \bibinfo {author} {\bibfnamefont {T.}~\bibnamefont {Enss}},\ }\href
  {\doibase 10.1103/PhysRevLett.112.135302} {\bibfield  {journal} {\bibinfo
  {journal} {Phys. Rev. Lett.}\ }\textbf {\bibinfo {volume} {112}},\ \bibinfo
  {pages} {135302} (\bibinfo {year} {2014})}\BibitemShut {NoStop}%
\bibitem [{\citenamefont {Fischer}\ and\ \citenamefont
  {Parish}(2014)}]{Parish2}%
  \BibitemOpen
  \bibfield  {author} {\bibinfo {author} {\bibfnamefont {A.~M.}\ \bibnamefont
  {Fischer}}\ and\ \bibinfo {author} {\bibfnamefont {M.~M.}\ \bibnamefont
  {Parish}},\ }\href {\doibase 10.1103/PhysRevB.90.214503} {\bibfield
  {journal} {\bibinfo  {journal} {Phys. Rev. B}\ }\textbf {\bibinfo {volume}
  {90}},\ \bibinfo {pages} {214503} (\bibinfo {year} {2014})}\BibitemShut
  {NoStop}%
\bibitem [{\citenamefont {Bertaina}\ and\ \citenamefont
  {Giorgini}(2011)}]{Bertaina2011}%
  \BibitemOpen
  \bibfield  {author} {\bibinfo {author} {\bibfnamefont {G.}~\bibnamefont
  {Bertaina}}\ and\ \bibinfo {author} {\bibfnamefont {S.}~\bibnamefont
  {Giorgini}},\ }\href {\doibase 10.1103/PhysRevLett.106.110403} {\bibfield
  {journal} {\bibinfo  {journal} {Phys. Rev. Lett.}\ }\textbf {\bibinfo
  {volume} {106}},\ \bibinfo {pages} {1} (\bibinfo {year} {2011})},\ \Eprint
  {http://arxiv.org/abs/1011.3737} {1011.3737} \BibitemShut {NoStop}%
\bibitem [{\citenamefont {Randeria}\ \emph {et~al.}(1990)\citenamefont
  {Randeria}, \citenamefont {Duan},\ and\ \citenamefont {Shieh}}]{Randeria2d}%
  \BibitemOpen
  \bibfield  {author} {\bibinfo {author} {\bibfnamefont {M.}~\bibnamefont
  {Randeria}}, \bibinfo {author} {\bibfnamefont {J.-M.}\ \bibnamefont {Duan}},
  \ and\ \bibinfo {author} {\bibfnamefont {L.-Y.}\ \bibnamefont {Shieh}},\
  }\href {\doibase 10.1103/PhysRevB.41.327} {\bibfield  {journal} {\bibinfo
  {journal} {Phys. Rev. B}\ }\textbf {\bibinfo {volume} {41}},\ \bibinfo
  {pages} {327} (\bibinfo {year} {1990})}\BibitemShut {NoStop}%
\bibitem [{\citenamefont {Nelson}\ and\ \citenamefont
  {Kosterlitz}(1977)}]{NelsonKosterlitz}%
  \BibitemOpen
  \bibfield  {author} {\bibinfo {author} {\bibfnamefont {D.~R.}\ \bibnamefont
  {Nelson}}\ and\ \bibinfo {author} {\bibfnamefont {J.~M.}\ \bibnamefont
  {Kosterlitz}},\ }\href {\doibase 10.1103/PhysRevLett.39.1201} {\bibfield
  {journal} {\bibinfo  {journal} {Phys. Rev. Lett.}\ }\textbf {\bibinfo
  {volume} {39}},\ \bibinfo {pages} {1201} (\bibinfo {year}
  {1977})}\BibitemShut {NoStop}%
\bibitem [{\citenamefont {Holzmann}\ \emph {et~al.}(2007)\citenamefont
  {Holzmann}, \citenamefont {Baym}, \citenamefont {Blaizot},\ and\
  \citenamefont {Lalo\"{e}}}]{Baym}%
  \BibitemOpen
  \bibfield  {author} {\bibinfo {author} {\bibfnamefont {M.}~\bibnamefont
  {Holzmann}}, \bibinfo {author} {\bibfnamefont {G.}~\bibnamefont {Baym}},
  \bibinfo {author} {\bibfnamefont {J.-P.}\ \bibnamefont {Blaizot}}, \ and\
  \bibinfo {author} {\bibfnamefont {F.}~\bibnamefont {Lalo\"{e}}},\ }\href
  {\doibase 10.1073/pnas.0609957104} {\bibfield  {journal} {\bibinfo  {journal}
  {Proc. Natl. Acad. Sci.}\ }\textbf {\bibinfo {volume} {104}},\ \bibinfo
  {pages} {1476} (\bibinfo {year} {2007})}\BibitemShut {NoStop}%
\bibitem [{\citenamefont {Dupuis}(2014)}]{Dupuis}%
  \BibitemOpen
  \bibfield  {author} {\bibinfo {author} {\bibfnamefont {N.}~\bibnamefont
  {Dupuis}},\ }\href {\doibase 10.1103/PhysRevB.89.035113} {\bibfield
  {journal} {\bibinfo  {journal} {Phys. Rev. B}\ }\textbf {\bibinfo {volume}
  {89}},\ \bibinfo {pages} {035113} (\bibinfo {year} {2014})}\BibitemShut
  {NoStop}%
\bibitem [{\citenamefont {Haussmann}\ \emph {et~al.}(2007)\citenamefont
  {Haussmann}, \citenamefont {Rantner}, \citenamefont {Cerrito},\ and\
  \citenamefont {Zwerger}}]{Zwerger}%
  \BibitemOpen
  \bibfield  {author} {\bibinfo {author} {\bibfnamefont {R.}~\bibnamefont
  {Haussmann}}, \bibinfo {author} {\bibfnamefont {W.}~\bibnamefont {Rantner}},
  \bibinfo {author} {\bibfnamefont {S.}~\bibnamefont {Cerrito}}, \ and\
  \bibinfo {author} {\bibfnamefont {W.}~\bibnamefont {Zwerger}},\ }\href
  {\doibase 10.1103/PhysRevA.75.023610} {\bibfield  {journal} {\bibinfo
  {journal} {Phys. Rev. A}\ }\textbf {\bibinfo {volume} {75}},\ \bibinfo
  {pages} {023610} (\bibinfo {year} {2007})}\BibitemShut {NoStop}%
\bibitem [{\citenamefont {Chen}\ \emph {et~al.}(1999)\citenamefont {Chen},
  \citenamefont {Kosztin}, \citenamefont {Jank\'o},\ and\ \citenamefont
  {Levin}}]{Chen1}%
  \BibitemOpen
  \bibfield  {author} {\bibinfo {author} {\bibfnamefont {Q.~J.}\ \bibnamefont
  {Chen}}, \bibinfo {author} {\bibfnamefont {I.}~\bibnamefont {Kosztin}},
  \bibinfo {author} {\bibfnamefont {B.}~\bibnamefont {Jank\'o}}, \ and\
  \bibinfo {author} {\bibfnamefont {K.}~\bibnamefont {Levin}},\ }\href@noop {}
  {\bibfield  {journal} {\bibinfo  {journal} {Phys. Rev. B}\ }\textbf {\bibinfo
  {volume} {59}},\ \bibinfo {pages} {7083} (\bibinfo {year}
  {1999})}\BibitemShut {NoStop}%
\bibitem [{\citenamefont {Chen}\ \emph {et~al.}(2005)\citenamefont {Chen},
  \citenamefont {Stajic}, \citenamefont {Tan},\ and\ \citenamefont
  {Levin}}]{ourreview}%
  \BibitemOpen
  \bibfield  {author} {\bibinfo {author} {\bibfnamefont {Q.~J.}\ \bibnamefont
  {Chen}}, \bibinfo {author} {\bibfnamefont {J.}~\bibnamefont {Stajic}},
  \bibinfo {author} {\bibfnamefont {S.~N.}\ \bibnamefont {Tan}}, \ and\
  \bibinfo {author} {\bibfnamefont {K.}~\bibnamefont {Levin}},\ }\href@noop {}
  {\bibfield  {journal} {\bibinfo  {journal} {Phys. Rep.}\ }\textbf {\bibinfo
  {volume} {412}},\ \bibinfo {pages} {1} (\bibinfo {year} {2005})}\BibitemShut
  {NoStop}%
\bibitem [{\citenamefont {Petrov}\ and\ \citenamefont
  {Shlyapnikov}(2001)}]{Petrov2000}%
  \BibitemOpen
  \bibfield  {author} {\bibinfo {author} {\bibfnamefont {D.~S.}\ \bibnamefont
  {Petrov}}\ and\ \bibinfo {author} {\bibfnamefont {G.~V.}\ \bibnamefont
  {Shlyapnikov}},\ }\href {\doibase 10.1103/PhysRevA.64.012706} {\bibfield
  {journal} {\bibinfo  {journal} {Phys. Rev. A}\ }\textbf {\bibinfo {volume}
  {64}},\ \bibinfo {pages} {012706} (\bibinfo {year} {2001})}\BibitemShut
  {NoStop}%
\bibitem [{\citenamefont {Hadzibabic}\ and\ \citenamefont
  {Dalibard}()}]{DalibardReview}%
  \BibitemOpen
  \bibfield  {author} {\bibinfo {author} {\bibfnamefont {Z.}~\bibnamefont
  {Hadzibabic}}\ and\ \bibinfo {author} {\bibfnamefont {J.}~\bibnamefont
  {Dalibard}},\ }\enquote {\bibinfo {title} {{BKT} physics with two-dimensional
  atomic gases},}\ in\ \href {\doibase 10.1142/9789814417648_0009} {\emph
  {\bibinfo {booktitle} {40 Years of Berezinskii--Kosterlitz--Thouless
  Theory}}},\ Chap.~\bibinfo {chapter} {9}, pp.\ \bibinfo {pages}
  {297--323}\BibitemShut {NoStop}%
\bibitem [{\citenamefont {Prokof'ev}\ and\ \citenamefont
  {Svistunov}(2002)}]{MonteCarlo}%
  \BibitemOpen
  \bibfield  {author} {\bibinfo {author} {\bibfnamefont {N.}~\bibnamefont
  {Prokof'ev}}\ and\ \bibinfo {author} {\bibfnamefont {B.}~\bibnamefont
  {Svistunov}},\ }\href {\doibase 10.1103/PhysRevA.66.043608} {\bibfield
  {journal} {\bibinfo  {journal} {Phys. Rev. A}\ }\textbf {\bibinfo {volume}
  {66}},\ \bibinfo {pages} {043608} (\bibinfo {year} {2002})}\BibitemShut
  {NoStop}%
\bibitem [{fit()}]{fitwords}%
  \BibitemOpen
  \href@noop {} {}\bibinfo {note} {We thus address the $r\rightarrow\infty$
  asymptotic form of $g_1(r)$ which in the homogenous limit gives $g_1(r) \sim
  e^{-r/\xi} r^{-1/2}$, where $\xi^2 \propto \lambda^2 e^{\mathcal{D}}$. At low
  temperatures, we see that the screening length $\xi\rightarrow \infty$, and
  we expect there is a range in $r$ well described by a power law with $\eta
  \sim 1/2$. If we include this exponential screening effect, and fit to the
  form $g_1(r) \sim e^{-r/\xi} r^{-\eta}$, we find our fits are significantly
  improved as found through a reduced chi-squared test.}\BibitemShut {Stop}%
\end{thebibliography}%

\clearpage


\section*{Supplementary material}

# Supplementary Material: Quasi-condensation in two-dimensional Fermi gases

Chien-Te Wu, Brandon M. Anderson, Rufus Boyack, and K. Levin
*James Franck Institute, University of Chicago, Chicago, Illinois 60637, USA*

## I. OVERVIEW OF THE PRESENT PSEUDOGAP FORMALISM

The theoretical formalism in this paper is a $t$-matrix theory chosen to reproduce the physics of a generalized BCS phase. This approach establishes the nature of non-condensed pairs of fermions, which will self-consistently generalize BCS theory. We confine ourselves to two dimensions where, as we have seen in the main text, there is no condensate. All fermion pairs are thus non-condensed. The BCS dispersion relation can be written as $E_\mathbf{k} \equiv \sqrt{\xi_\mathbf{k}^2 + \Delta^2}$, where we write the gap parameter as $\Delta$ and make it clear that this represents a pairing gap and not an order parameter. In all $t$-matrix theories the $t$-matrix, $\Gamma(Q)$, is written as

$$\Gamma(Q) = \frac{g}{1 + g\chi(Q)}. \tag{S1}$$

Each $t$-matrix theory differs only in the choice of the pair susceptibility, $\chi(Q)$.

To make the appropriate choice, here we note that the mean-field gap equation can be expressed as $\Gamma^{-1}(0) = 0$, where

$$\Gamma^{-1}(0) = \sum_\mathbf{k} \left[ \frac{1 - f(E_\mathbf{k}) - f(\xi_\mathbf{k})}{E_\mathbf{k} + \xi_\mathbf{k}} u_\mathbf{k}^2 - \frac{f(E_\mathbf{k}) - f(\xi_\mathbf{k})}{E_\mathbf{k} - \xi_\mathbf{k}} v_\mathbf{k}^2 \right] + g^{-1}. \tag{S2}$$

Note that this is associated with a pair of Green's functions, one of which is dressed and one of which is bare. We now write a natural generalization of $\Gamma(0)$ to $\Gamma(Q)$. This leads to

$$\Gamma^{-1}(Q) = \sum_\mathbf{k} \left[ \frac{1 - f(E_\mathbf{k}) - f(\xi_{\mathbf{k}-\mathbf{q}})}{E_\mathbf{k} + \xi_{\mathbf{k}-\mathbf{q}} - \Omega - i0^+} u_\mathbf{k}^2 - \frac{f(E_\mathbf{k}) - f(\xi_{\mathbf{k}-\mathbf{q}})}{E_\mathbf{k} - \xi_{\mathbf{k}-\mathbf{q}} + \Omega + i0^+} v_\mathbf{k}^2 \right] + g^{-1}, \tag{S3}$$

which corresponds to $\chi(Q) = \sum_K G(K) G_0(Q - K)$ where $G_0(K)$ is a non-interacting Green's function, and the full Green's function $G(K)$ includes a self-energy $\Sigma(K)$ through $G^{-1}(K) = G_0^{-1}(K) - \Sigma(K)$. The specific form of Eq. (S3) requires a BCS-like self-energy of the form $\Sigma(K) = -\Delta^2 G_0(-K)$. Note that the general $GG_0$ form of $\chi$ can be derived using the equations of motion for the four-point correlation function in the ladder diagram approximation and is known to be the only $t$-matrix consistent with the BCS gap equation [S1, S2].

We now show that the above $t$-matrix theory is consistent with a BCS-like pairing gap in the limit that $\mu_\text{pair}$ is sufficiently small. As in all $t$-matrix theories, the fermions acquire a self-energy which is given by

$$\Sigma(K) = \sum_Q \Gamma(Q) G_0(Q - K). \tag{S4}$$

If the pairs are close to condensation (with $\Gamma^{-1}(0) \approx 0$, so that $\Gamma(0) \approx \infty$) then we can approximate the above equation by

$$\Sigma(K) \approx \left( \sum_Q \Gamma(Q) \right) G_0(-K) = -\Delta^2 G_0(-K), \tag{S5}$$

where in the second equality we have defined $\Delta^2 \equiv -\sum_Q \Gamma(Q)$ to show that this approximation is consistent with a BCS-like self-energy. Note that $\Delta^2$ enters only as a pairing gap in the Bogoliubov spectrum $E_\mathbf{k}$, and does not need to represent an order parameter in any way. Thus, we have now established a self-consistent $t$-matrix formalism that reproduces the BCS gap equation in the limit that $\mu_\text{pair} = 0$.

At small $Q$, the $t$-matrix can be expanded in the form

$$\Gamma(Q) = \frac{a_0^{-1}}{\Omega - \Omega_\mathbf{q} + \mu_\text{pair} + i\gamma_Q}, \tag{S6}$$

where the pair chemical potential is $\mu_{\text{pair}} = a_0^{-1} \left( g^{-1} + \chi(0) \right)$ and we introduce the bosonic effective mass from a small momentum pair dispersion $\Omega_{\mathbf{q}} \approx q^2/2M_B$. The specific expressions for $M_B$ and $\gamma_Q$ are complicated and are the same as those in Refs. [S1, S2] up to a replacement of a 3D sum with a 2D sum. The constant $a_0 = (n - n_0)/2\Delta^2$, where $n = 2\sum_K G(K)$ and $n_0 = 2\sum_K G_0(K)$. From our definition of $\Delta^2$, after summation over bosonic Matsubara frequencies

$$\Delta^2 = -\sum_Q \Gamma(Q) = \frac{1}{a_0} \sum_{\mathbf{q}} b(\Omega_{\mathbf{q}} - \mu_{\text{pair}}), \tag{S7}$$

where $b(x) = (e^{x/k_B T} - 1)^{-1}$ is the Bose-Einstein distribution function as defined in the main text. This allows us to identify

$$n_B = a_0 \Delta^2 = \frac{1}{2}(n - n_0), \tag{S8}$$

where $n_B$ represents the total number of non-condensed bosons.

In summary, the essential approximation in this scheme is the assumption that $\mu_{\text{pair}}$ is small, so that Eq. (S5) is valid. As a consequence, the fermionic self-energy is of the BCS-type, and the number equation is essentially that of BCS theory with the pairing gap $\Delta$.

These arguments lead to the three coupled equations in Eqs. (2–4) of the main text. However, we can use the definition of $\mu_{\text{pair}}$ in the text to reduce the system of equations to be solved for to:

$$\sum_{\mathbf{k}} \left[ \frac{1 - 2f(E_{\mathbf{k}})}{2E_{\mathbf{k}}} - \frac{1}{2\epsilon_{\mathbf{k}} + \epsilon_B} \right] = a_0 T \ln \left( 1 - e^{-n_B \lambda_B^2} \right), \tag{S9}$$

$$\sum_{\mathbf{k}} \left[ 1 - \frac{\xi_{\mathbf{k}}}{E_{\mathbf{k}}} (1 - 2f(E_{\mathbf{k}})) \right] = n. \tag{S10}$$

This is now two equations in two unknowns. Here $\epsilon_B$ is the binding energy defined in the main text in terms of the scattering length and $\lambda_B = \sqrt{2\pi\hbar^2/M_B k_B T}$ is the thermal wavelength for the bosonic pairs.

## II. TRAP EFFECTS

In the presence of a trap, the homogeneous mean-field equations are no longer valid. However, some trap effects can be included if we rewrite our equations using the transformations $\mu \to \mu(\mathbf{R}) = \mu_0 - \frac{1}{2} m\omega^2 \mathbf{R}^2$, and $\Delta \to \Delta(\mathbf{R})$, where $\mathbf{R}$ is a local position, and is not to be confused with the conjugate to $\mathbf{q}$. Note that $\mu_0$ is still homogeneous in space (i.e., there is only one degree of freedom) but that $\Delta(\mathbf{R})$ is no longer homogeneous. We now have $\xi_{\mathbf{k}} \to \xi_{\mathbf{k}}(\mathbf{R}) = k^2/2m - \mu(\mathbf{R})$, and $E_{\mathbf{k}} \to E_{\mathbf{k}}(\mathbf{R}) = \sqrt{\xi_{\mathbf{k}}^2(\mathbf{R}) + \Delta^2(\mathbf{R})}$. Similarly, $M_B \to M_B(\mathbf{R})$, $n_B \to n_B(\mathbf{R})$, $a_0 \to a_0(\mathbf{R})$, etc. through these same substitutions.

Using these substitutions, known as the local density approximation (LDA), the mean-field equations now become

$$\sum_{\mathbf{k}} \left[ \frac{1 - 2f(E_{\mathbf{k}}(R))}{2E_{\mathbf{k}}(R)} - \frac{1}{2\epsilon_{\mathbf{k}} + \epsilon_B} \right] = \frac{n_B(R)}{\Delta^2(R)} T \ln \left( 1 - e^{-n_B(R)\lambda_B^2(R)} \right), \tag{S11}$$

$$2\pi \int dR\, R \left[ \sum_{\mathbf{k}} 1 - \frac{\xi_{\mathbf{k}}(R)}{E_{\mathbf{k}}(R)} (1 - 2f(E_{\mathbf{k}}(R))) \right] = N, \tag{S12}$$

where $N$ is the total number of particles, and we have used cylindrical symmetry. These equations are discretized on a grid of $M$ points in $R$ space. Equation (S11) is split into $M$ equations, whereas Eq. (S12) is a single equation. Therefore, we have $M + 1$ total equations. Furthermore, since we have discretized $\Delta(R)$ into $M$ points with $\mu_0$ a single parameter, we have $M + 1$ degrees of freedom. Thus, there are an equal number of constraints as there are degrees of freedom.

Within the LDA, the solutions to the mean-field gap equations at $T=0$ can be written analytically:

$$\Delta(R) = \sqrt{2\epsilon_B \epsilon_F} \sqrt{1 - \left(\frac{R}{R_{TF}}\right)^2} \Theta\left(1 - \left(\frac{R}{R_{TF}}\right)^2\right), \tag{S13}$$

$$\mu(R) = \epsilon_F \left(1 - \left(\frac{R}{R_{TF}}\right)^2\right) - \frac{\epsilon_B}{2}. \tag{S14}$$

Here the Thomas-Fermi radius $R_{TF}$ is defined by $m\omega^2 R_{TF}^2/2 = \epsilon_F$.

In the LDA, the Fermi energy is defined from the $T = 0$ and $\Delta = 0$ limit which, including the factor of two for spin degeneracy, gives $\epsilon_F = \sqrt{N}\hbar\omega$. From the LDA chemical potential $\mu(0) = \epsilon_F$, and we can connect the central density, $n(0)$, to the Fermi energy in the usual way: $\epsilon_F = \hbar^2 k_F^2/2m$, where $k_F^2 = 2\pi n(0)$ defines the Fermi momentum $\hbar k_F$. Throughout our numerical calculations we use $N = 200^2$ particles, consistent with the experimental value of $N = 40,000$ total atoms. This relation then implies $\hbar\omega/\epsilon_F = 1/200$.

In both the theory and the experiment, the scattering parameter is input as $\ln(k_F a_{2D})$. Note that the experiment estimates the Fermi energy through the central density at finite temperature leading to a slightly different value of $\ln(k_F a_{2D})$ for an equivalent $a_{2D}$. To show this deviation is not significant, we define $r = \sqrt{N}\hbar\omega/(\pi\hbar^2 n(0)/m)$, i.e., the ratio of the Fermi energy as defined from the trapping frequency and atom number, to the Fermi energy as defined from the central density. Using the numbers from Ref. [S3], we find $r$ is in a range of $r \sim 0.25$ to $r \sim 0.9$ in the BEC and BCS limit respectively. Thus there may be a weak difference in comparing our scattering length parameter, $\ln(k_F a_{2D})$, relative to the experiment [S3, S4]. We do not expect this correction to significantly modify our analysis or comparison with experiment.

### III. COMMENTS ON OTHER THEORETICAL APPROACHES

Alternative theories of 2D superfluids at finite temperature (which treat BKT physics to varying degrees) fall into two main classes: (1) Theories which have a BCS basis and (using path integral techniques [S5–S9]) focus on determining the phase stiffness parameter, $\rho_s$. This is then substituted into the BKT transition temperature expression. (2) Those which adopt one of two $t$-matrix approaches [S10–S13] neither of which, in contrast to the present paper, is based on BCS theory.

The first class of theories [S6–S9], as noted in Ref. [S7], neglects non-condensed pair degrees of freedom [S14], in contrast to the second class [S10–S13], but in the first class, BKT physics plays a more prominent role. All homogeneous theories in 2D must satisfy the Mermin-Wagner theorem. For the first class of theories this theorem can be directly imposed by performing an average over phase fluctuations. As for the second class, we quote from Ref. [S7], that the Mermin-Wagner theorem "*is an excellent check on whether one has a good solution to the t-matrix equations: the correct solution should give $T_c = 0$ in 2D.*" We have verified this consistency in the present $t$-matrix formalism.

In addition, there are theories which introduce inhomogeneity, trap and/or finite size effects [S5, S15] which thereby enable the calculation of a finite transition temperature often also associated with the BKT phase and there is also an extensive literature on the ground state [S16, S17].

As for alternative $t$-matrix theories, in Ref. [S13] a Baym-Luttinger-Ward approach is applied to study a 2D homogeneous Fermi gas. In the Baym-Luttinger-Ward formalism the self-energy arises as a functional derivative of the Luttinger-Ward functional with respect to the full Green's function. The advantage of this formalism is that it is manifestly conserving and self-consistent. However, if one truncates this infinite system of diagrams, then the theory is no longer conserving. Thus, in 3D for example, satisfying the Thouless criterion requires a redefinition of the coupling constant parameters [S18]. In 2D, where there is no Thouless criterion, one must regularize the theory to ensure the Mermin-Wagner theorem is satisfied [S19]. Reference [S13] applies a condition that $\Gamma^{-1}(0) \propto 1/N \neq 0$, which appears in the BEC literature [S20], and sets $N = 500$. Due to numerical difficulties they are only able to address this BKT transition on the BCS side of resonance.

Recent work [S11] based on a $G_0 G_0$ $t$-matrix formalism has presented results for the pair momentum distribution which appear to have some overlap with the present paper and the experimental data in Ref. [S3]. However, in contrast to both this paper and experiment, Ref. [S11] does not include a trap, which makes a comparison between theory and experiment difficult. Indeed, in Ref. [S12] the authors examined the effects of a trap within the same theory and found a very high superfluid transition temperature (not related to BKT physics).

---

[S1] Q. J. Chen, I. Kosztin, B. Jankó, and K. Levin, Phys. Rev. B **59**, 7083 (1999).
[S2] Q. J. Chen, J. Stajic, S. N. Tan, and K. Levin, Phys. Rep. **412**, 1 (2005).
[S3] M. G. Ries, A. N. Wenz, G. Zürn, L. Bayha, I. Boettcher, D. Kedar, P. A. Murthy, M. Neidig, T. Lompe, and S. Jochim, Phys. Rev. Lett. **114**, 230401 (2015).
[S4] P. A. Murthy, I. Boettcher, L. Bayha, M. Holzmann, D. Kedar, M. Neidig, M. G. Ries, A. N. Wenz, G. Zürn, and S. Jochim, Phys. Rev. Lett. **115**, 10401 (2015).
[S5] D. S. Petrov, M. A. Baranov, and G. V. Shlyapnikov, Phys. Rev. A **67**, 031601 (2003).
[S6] E. Babaev and H. Kleinert, Phys. Rev. B **59**, 12083 (1999).

[S7] V. M. Loktev, R. M. Quick, and S. G. Sharapov, Physics Reports **349**, 1 (2001).
[S8] S. S. Botelho and C. A. R. Sá de Melo, Phys. Rev. Lett. **96**, 040404 (2006).
[S9] L. Salasnich, P. A. Marchetti, and F. Toigo, Phys. Rev. A **88**, 053612 (2013).
[S10] F. Marsiglio, P. Pieri, A. Perali, F. Palestini, and G. C. Strinati, Phys. Rev. B **91**, 054509 (2015).
[S11] M. Matsumoto, D. Inotani, and Y. Ohashi, (2015), arXiv:1507.05149.
[S12] R. Watanabe, S. Tsuchiya, and Y. Ohashi, Phys. Rev. A **88**, 013637 (2013).
[S13] M. Bauer, M. M. Parish, and T. Enss, Phys. Rev. Lett. **112**, 135302 (2014).
[S14] Recent work by B. Bighin and L. Salasnich (arXiv 1507.97542) does introduce bosonic degrees of freedom into a path integral scheme via the gapless Gaussian fluctuations. However, it has been argued (by C. C. Chien et al, Annals of Physics 347, 192 (2014)) that there should not be gapless Goldstone bosons in 2D systems.
[S15] A. M. Fischer and M. M. Parish, Phys. Rev. B **90**, 214503 (2014).
[S16] G. Bertaina and S. Giorgini, Phys. Rev. Lett. **106**, 1 (2011), 1011.3737.
[S17] M. Randeria, J.-M. Duan, and L.-Y. Shieh, Phys. Rev. B **41**, 327 (1990).
[S18] R. Haussmann, W. Rantner, S. Cerrito, and W. Zwerger, Phys. Rev. A **75**, 023610 (2007).
[S19] N. Dupuis, Phys. Rev. B **89**, 035113 (2014).
[S20] M. Holzmann, G. Baym, J.-P. Blaizot, and F. Laloë, Proc. Natl. Acad. Sci. **104**, 1476 (2007).



\end{document}